\documentclass{kluwer}

\usepackage{graphicx}

\begin{document}

\newcommand{\lya}{Lyman~$\alpha$}
\newcommand{\lyb}{Lyman~$\beta$}
\newcommand{\za}{$z_{\rm abs}$}
\newcommand{\ze}{$z_{\rm em}$}
\newcommand{\cmtwo}{cm$^{-2}$}
\newcommand{\nhi}{$N$(H$^0$)}
\newcommand{\degpoint}{\mbox{$^\circ\mskip-7.0mu.\,$}}
\newcommand{\halpha}{\mbox{H$\alpha$}}
\newcommand{\hbeta}{\mbox{H$\beta$}}
\newcommand{\hgamma}{\mbox{H$\gamma$}}
\newcommand{\kms}{\,km~s$^{-1}$}      
\newcommand{\minpoint}{\mbox{$'\mskip-4.7mu.\mskip0.8mu$}}
\newcommand{\mv}{\mbox{$m_{_V}$}}
\newcommand{\Mv}{\mbox{$M_{_V}$}}
\newcommand{\secpoint}{\mbox{$''\mskip-7.6mu.\,$}}
\newcommand{\sqdeg}{\mbox{${\rm deg}^2$}}
\newcommand{\squig}{\sim\!\!}
\newcommand{\subsun}{\mbox{$_{\twelvesy\odot}$}}
\newcommand{\et}{et al.~}

\def\ltsima{$\; \buildrel < \over \sim \;$}
\def\simlt{\lower.5ex\hbox{\ltsima}}
\def\gtsima{$\; \buildrel > \over \sim \;$}
\def\simgt{\lower.5ex\hbox{\gtsima}}
\def\arcs{$''~$}
\def\arcm{$'~$}

\begin{article}
\begin{opening}
\title{The Evolution and Space Density of Damped Lyman-$\alpha$
Galaxies}

\author{C\'{e}line \surname{P\'{e}roux}\thanks{e-mail: celine@ast.cam.ac.uk}}
\author{Mike J. \surname{Irwin}} 
\author{Richard G. \surname{McMahon}}
\institute{Institute of Astronomy, University of Cambridge, UK}                               
\author{Lisa J. \surname{Storrie-Lombardi}}
\institute{SIRTF Science Center, California Institute of Technology, Pasadena, USA}




\runningtitle{The Evolution and Space Density of Damped Lyman-$\alpha$
Galaxies}
\runningauthor{P\'{e}roux et al.}



\begin{abstract} 
The results of a new spectroscopic survey of 66 $z\simgt4$ quasars for
Damped Lyman-$\alpha$ absorption systems are presented. The search led
to the discovery of 30 new DLA candidates which are analysed in order
to compute the comoving mass density of neutral gas in a non-zero
$\Lambda$ Universe. The possible sources of uncertainty are discussed
and the implications of our results for the theories of galaxy
formation and evolution are emphasized. A subsequent paper will
present details of the calculations summarised here and a more
extensive explanation of the consequences of our observations for the
understanding of the nature of DLAs.
\end{abstract}


\end{opening}

\section{A New Survey for Damped Lyman-$\alpha$ Galaxies}

Damped Lyman-$\alpha$ (hereafter DLA) galaxies are defined as
absorption systems in the spectra of quasars with a hydrogen column
density $N(HI)>2\times 10^{20}$ atoms cm$^{-2}$ ($\geq10$ \AA\space
rest frame equivalent width, Wolfe et al. 1986). Their space density
distribution, metal abundances and kinematics seem to indicate that
they are the progenitors of present day galaxies, thus providing a
powerful observational tool for galaxy evolution studies.
\newline
DLAs probe the neutral gas from which stars form and thus are a direct
means to trace galaxy formation. Quasars are extremely luminous
objects which are observed up to $z=5.8$ (Fan et al. 2000), and thus
DLAs can be probed over a large redshift range. Finally DLA galaxies
can be detected to fainter limits than objects detected using
traditional emission observations.
\newline
We observed 66 $z\simgt4$ quasars (see Peroux et al. 2000 for details
on the observations) from the $2^{nd}$ APM colour survey
(Storrie-Lombardi et al. 2000), the Palomar Sky Survey (Kennefick et
al. 1995a $\&$ 1995b and G. Djorgovski's web page) and the Sloan
Digital Sky Survey (Fan et al. 1999). This resulted in 31 DLA
candidates, 30 of which are newly discovered and many with associated
metal lines.

\section{Probing the Neutral Gas Content of the Universe}

\begin{figure}
\centerline{\includegraphics[height=7cm,
width=9cm,angle=0]{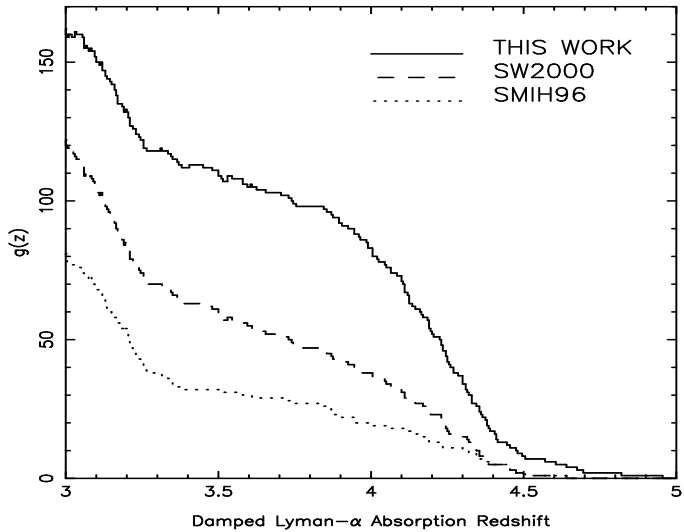}}
\caption{Sensitivity functions for DLA galaxy searches. The dotted
line is from Storrie-Lombardi et al. 1996 ($\Delta(z)=239$), the
dashed line is the compilation of Storrie-Lombardi $\&$ Wolfe 2000
($\Delta(z)=418$), and our survey is represented by the solid line
($\Delta(z)=493$). Our new sample more than doubles the redshift path
surveyed above $z\simgt3.5$.}
\end{figure}

For each quasar the redshift range along which a DLA could be observed
is determined and used to deduce the survey's sensitivity. This is
represented by the $g(z)$ function shown in Fig 1. Our new data is
analysed in conjunction with previous observations resulting in 115
DLAs observed in 697 quasars (Rao and Turnshek 1999 data not
included).
\newline
The mass of neutral gas contained in DLAs is expressed as:
$\Omega_{DLA}(z) = \frac{H_o \mu m_H}{c \rho_{crit}}
\int_{N_{min}}^{\infty} N f(N,z) dN$, where $N$ is the column density
and $f(N,z)$ is the distribution function. The integral is estimated
by summing up over all the observed column densities and dividing by
the absorption distance interval, $X(z)$, which corrects for comoving
coordinates and thus depends on the geometry of the Universe. In a
non-zero $\Lambda$ Universe: $X(z)=\int_{0}^{z}(1+z)^2 \left[ (1+z)^2
( 1 + z \Omega_M) - z ( 2 + z ) \Omega_{\Lambda} \right]^{-1/2} dz$.

\begin{figure}
\centerline{\includegraphics[height=9cm, width=11.5cm,angle=0]{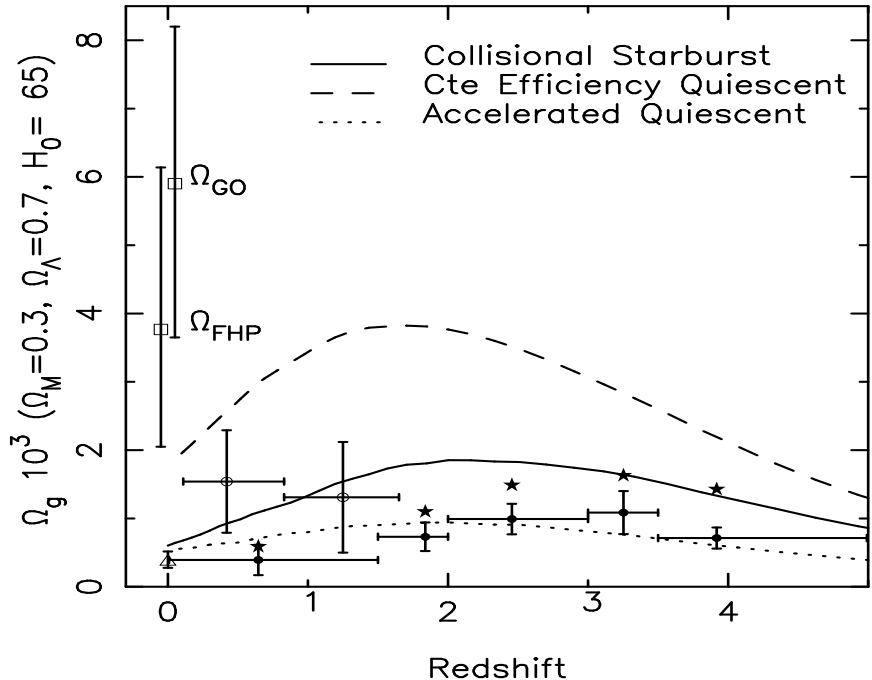}} 
\caption{The circles show the neutral gas in Damped Lyman-$\alpha$
galaxies (the filled circles are our observations and the open circles
are Rao $\&$ Turnshek 1999 results). Vertical bars correspond to 1
$\sigma$ errors and the horizontal bars indicate bin sizes. The stars
are the total HI including a correction for the neutral gas contained
in systems with column densities below $2\times 10^{20}$ atoms
cm$^{-2}$. The triangle at $z=0$ is from Natarajan $\&$ Pettini 1997
who used a recent galaxy luminosity function to reproduce the local HI
mass measured by Rao $\&$ Turnshek 1993. The squares, $\Omega_{FHP}$
and $\Omega_{GO}$ (Fukugita, Hogan $\&$ Peebles 1998 and Gnedin $\&$
Ostriker 1992 respectively) are $\Omega_{baryons}$ in local
galaxies. The semi-analytical models from Somerville, Primack $\&$
Faber 2000 are overplotted (see text for detailed explanations of the
models).}
\end{figure}

Our observations are presented in Fig 2 for a currently popular
cosmology and show the evolution of $\Omega_{DLA}$ over the last 90
$\%$ of the age of the Universe thus probing the early stages of
galaxies even before gas turns into stars.
\newline
In order to account for the neutral gas not included in DLAs, we used
a power law with an exponential turnover (similar to the Schechter
function) in extrapolating from the observed column density
distribution: $f(N,z)=(f_*/N_*)(N/N_*)^{-\beta}e^{-N/N_*}$. Using the
parameters determined by Storrie-Lombardi et al. 1996,
$N_*=21.63\pm0.35$ and $\beta=1.48\pm0.30$, results in a $\sim20\%$
increase in $\Omega$ for $z\leq3.5$. For $z>3.5$ systems, our
observations indicate a steepening of the low column density end of
the DLA distribution (where much of the mass resides) in comparison
with our previous study and leads to a factor of $\sim2$ correction
(as shown by the stars in Fig 2).

\section{Discussion}

Most of the uncertainties discussed here seem to indicate that
$\Omega_{DLA}$ is being underestimated. At low redshift, Rao $\&$
Turnshek 1999 used a different method (based on the observations of
Lyman-$\alpha$ in identified Mg II systems) and HST data to derive a
higher neutral gas content in DLAs (Fig 2). These results are
surprisingly high and might be due to low number statistics or might
indicate that high redshift results are underestimated. More
uncertainties arise in the fact that dust is likely to produce
observational biases unaccounted for here and that the exact nature of
DLAs is unknown. In particular, they might be different types of
object at different redshift. Finally, higher resolution data are
necessary in order to identify multiple systems.
\newline
Several groups (Kauffmann $\&$ Haehnelt 2000, Somerville, Primack $\&$
Faber 2000) have included more realistic physics in their simulations
to construct semi-analytical models of galaxy formation which, among
other things, predict the evolution of cold gas in the Universe. The
models presented in Fig 2 vary in their recipe for star formation: it
is triggered by galaxy-galaxy mergers in the collisional starburst
model, constant with redshift in the constant efficiency model, and
scales inversely with disc dynamical time in the accelerated
efficiency model. Our observational results can thus be used to
directly constrain theories of galaxy evolution.
\newline
We would like to thank Isobel Hook for help in acquiring the data. CP
is grateful to Max Pettini and the organisers for help with a very
enjoyable conference.

{}

\end{article}
\end{document}